\documentclass[twocolumn,pra,aps,showpacs]{revtex4}
\usepackage{makeidx}
\usepackage{amsmath}
\usepackage{amssymb}
\usepackage{epsfig}
\usepackage[english]{babel}
\usepackage{latexsym}
\usepackage{graphics}
\usepackage{subfigure}

\begin{document}

\title{Transition to Instability in a Periodically Kicked Bose-Einstein
Condensate on a Ring}
\author{Jie Liu$^{1,2}$}
\thanks{These authors contributed equally to this work.}
\author{Chuanwei Zhang$^{2,3}$}
\thanks{These authors contributed equally to this work.}
\author{Mark G. Raizen$^{2,3}$}
\author{Qian Niu$^{2}$}

\begin{abstract}
A periodically kicked ring of a Bose-Einstein condensate is considered as a
nonlinear generalization of the quantum kicked rotor, where the nonlinearity
stems from the mean field interactions between the condensed atoms. For weak
interactions, periodic motion (anti-resonance) becomes quasiperiodic
(quantum beating) but remains stable. There exists a critical strength of
interactions beyond which quasiperiodic motion becomes chaotic, resulting in
an instability of the condensate manifested by exponential growth in the
number of noncondensed atoms. In the stable regime, the system remains
predominantly in the two lowest energy states and may be mapped onto a spin
model, from which we obtain an analytic expression for the beat frequency
and discuss the route to instability. We numerically explore parameter
regime for the occurrence of instability and reveal the characteristic
density profile for both condensed and noncondensed atoms. The Arnold
diffusion to higher energy levels is found to be responsible for the
transition to instability. Similar behavior is observed for dynamically
localized states (essentially quasiperiodic motions), where stability
remains for weak interactions but is destroyed by strong interactions.
\end{abstract}

\affiliation{$^1$Institute of Applied Physics and Computational Mathematics, Beijing,
100088, People's Republic of China \\
$^2$Department of Physics, The University of Texas, Austin, Texas
78712-1081, USA\\
$^3$Center for Nonlinear Dynamics, The University of Texas, Austin, Texas
78712-1081, USA
}
\pacs{03.75.-b, 05.45.-a, 03.65.Bz, 42.50.Vk }
\maketitle

\section{Introduction}

The $\delta $-kicked rotor is a textbook paradigm for the study of classical
and quantum chaos \cite{reichl}. In the classical regime, increasing kick
strengths destroy regular periodic or quasiperiodic motions of the rotor and
lead to the transition to chaotic motions, characterizing by diffusive
growth in the kinetic energy. In quantum mechanics, chaos is no longer
possible because of the linearity of the Schr\"{o}dinger equation and the
motion becomes periodic (anti-resonance), quasiperiodic (dynamical
localization), or resonant (quantum resonance) \cite{QKR,anti-res}.
Experimental study of these quantum phenomena have been done with ultra-cold
atoms in periodically pulsed optical lattices \cite{raizen}. However, most
of previous investigations have been focused on single particle systems and
the effects of interaction between particles have not received much
attention \cite{NLSE,zoller}.

In recently years, the realization of Bose-Einstein condensation (BEC) \cite%
{BEC} of dilute gases has opened new opportunities for studying dynamical
systems in the presence of many-body interactions. One can not only prepare
initial states with unprecedented precision and pureness, but also has the
freedom of introducing interactions between the particles in a controlled
manner. A natural question to ask is how the physics of the quantum kicked
rotor is modified by the interactions. In the mean field approximation,
many-body interactions in BEC are represented by adding a nonlinear term in
the Schr\"{o}dinger equation \cite{review} (such nonlinear Schr\"{o}dinger
equation also appears in the context of nonlinear optics \cite{nonoptics}).
This nonlinearity makes it possible to bring chaos back into the system,
leading to instability (in the sense of exponential sensitivity to initial
conditions) of the condensate wave function \cite{instability}. The onset of
instability of the condensate can cause rapid proliferation of thermal
particles \cite{castin} that can be observed in experiments \cite{ketterle}.
It is therefore important to understand the route to chaos with increasing
interactions. This problem has recently been studied for the kicked BEC in a
harmonic oscillator \cite{zoller}.

In Ref. \cite{zhang1}, we have investigated the quantum dynamics of a BEC
with repulsive interaction that is confined on a ring and kicked
periodically. This system is a nonlinear generalization of the quantum
kicked rotor. From the point of view of dynamical theory, the kicked rotor
is more generic than the kicked harmonic oscillator, because it is a typical
low dimensional system that obeys the KAM theorem, while the kicked harmonic
oscillator is known to be a special degenerate system out of the framework
of the KAM theorem \cite{liu1}. It is very interesting to understand how
both quantum mechanics and mean field interaction affect the dynamics of
such a generic system.

In this paper, we extend the results of Ref. \cite{zhang1}, including a more
detailed analysis of the model considered there as well as new phenonmena.
We will focus our attention on the relatively simpler case of quantum
anti-resonance, and show how the state is driven towards chaos or
instability by the mean field interaction. The paper is orgnaized as
follows: Section II lays out our physical model and its experimental
realization. Section III is devoted to the case of weak interactions between
atoms. We find that weak interactions make the periodic motion
(anti-resonance) quasiperiodic in the form of quantum beating. However, the
system remains predominantly in the lowest two energy levels and analytic
expressions for the beating frequencies are obtained by mapping the system
onto a spin model. Through varying the kick period, we find the phenomenon
of anti-resonance may be recovered even in the presence of interactions. The
decoherence effects due to thermal noise are discussed. Section IV is
devoted to the case of strong interactions. It is found that there exists a
critical strength of interactions beyond which quasiperiodic motion (quantum
beating) are destroyed, resulting in a transition to instability of the
condensate characterized by an exponential growth in the number of
noncondensed atoms. Universal critical behavior for the transition is found.
We show that the occurrence of instability corresponds to the process of
Arnold diffusion, through which the state can penetrate through the KAM tori
and escape to high energy levels \cite{Arnold}. We study nonlinear effects
on dynamically localized states that may be regarded as quasiperiodic \cite%
{hog}. Similar results are obtained in that localization remians for
sufficiently weak interactions but become unstable beyond a critical
strength of interactions. Section V consists of conclusions.

\section{Physical model: Kicked BEC on a ring}

Although the results obtained in this paper are common properties of systems
whose dynamics are governed by the nonlinear Schr\"{o}dinger equation, we
choose to present them here for a concrete physical model: a kicked BEC
confined on a ring trap, where the physical meanings of the results are easy
to understand. The description of the dynamics of this system may be divided
into two parts: condensed atoms and non-condensed atoms.

\subsection{Dynamics of condensed atoms: Gross-Pitaevskii equation}

Consider condensed atoms confined in a toroidal trap of radius $R$ and
thickness $r$, where $r\ll R$ so that lateral motion is negligible and the
system is essentially one-dimensional \cite{ring}. The dynamics of the BEC
is described by the dimensionless nonlinear Gross-Pitaveskii (GP) equation, 
\begin{equation}
i\frac{\partial }{\partial t}\psi =\left( -\frac{1}{2}\frac{\partial ^{2}}{%
\partial \theta ^{2}}+g\left\vert \psi \right\vert ^{2}+K\cos \theta \delta
_{t}(T)\right) \psi ,  \label{G-P}
\end{equation}%
where $g=8NaR/r^{2}$ is the scaled strength of nonlinear interaction, $N$ is
the number of atoms, $a$ is the $s$-wave scattering length, $K$ is the kick
strength, $\delta _{t}(T)$ represents $\sum\limits_{n}\delta \left(
t-nT\right) $, $T$ is the kick period, and $\theta $ denotes the azimuthal
angle. The length and the energy are measured in units $R$ and $\frac{\hbar
^{2}}{mR^{2}}$, respectively. The wavefunction $\psi \left( \theta ,t\right) 
$ has the normalization $\int_{0}^{2\pi }\left\vert \psi \right\vert
^{2}d\theta =1$ and satisfies periodic boundary condition $\psi \left(
\theta ,t\right) =\psi \left( \theta +2\pi ,t\right) $.

Experimentally, the ring-shape potential may be realized using two 2D
circular \textquotedblleft optical billiards" with the lateral dimension
being confined by two plane optical billiards \cite{billiard}, or
optical-dipole traps produced by red-detuned Laguerre-Gaussian laser beams
of varying azimuthal mode index \cite{ringexp}. The $\delta $-kick may be
realized by adding potential points along the ring with an off-resonant
laser \cite{raizen}, or by illuminating the BEC with a periodically pulsed
strongly detuned running wave laser in the lateral direction whose intensity 
$I$ is engineered to $I=I_{0}+I_{1}x/R$ \cite{Meck}. In the experiment, we
have the freedom to replace the periodic modulation $\cos \left( \theta
\right) $ of the kick potential with $\cos \left( j\theta \right) $, where $%
j $ is an integer. Such replacement revises the scaled interaction constant $%
g$ and kicked strength $K$, but will not affect the results obtained in the
paper. The interaction strength $g$ may be adjusted using a magnetic
field-dependent Feshbach resonance or the variation of the number of atoms 
\cite{Feshbach}.

\subsection{Dynamics of non-condensed atoms: Bogoliubov equation}

The deviation from the condensate wave function is described by Bogoliubov
equation that is obtained from a linearization around GP equation \cite%
{castin}. In Castin and Dum's formalism, the mean number of noncondensed
atoms at zero temperature is described by $\langle \delta \hat{N}(t)\rangle
=\sum_{k=1}^{\infty }\langle v_{k}(t)|v_{k}(t)\rangle $, where $v_{k}(t)$
are governed by 
\begin{equation}
i\frac{d}{dt}\left( \!%
\begin{array}{c}
u_{k} \\ 
v_{k}%
\end{array}%
\!\right) =\left( \!%
\begin{array}{cc}
H_{1} & H_{2} \\ 
-H_{2}^{\ast } & \!-H_{1}^{\ast }%
\end{array}%
\!\right) \!\left( \!%
\begin{array}{c}
u_{k} \\ 
v_{k}%
\end{array}%
\!\right) ,  \label{Bogo}
\end{equation}%
where $H_{1}=\hat{p}^{2}/2+g|\psi |^{2}-\mu +gQ|\psi |^{2}Q+K\cos \theta
\delta _{t}(T),$ $\!H_{2}=gQ\psi ^{2}Q^{\ast }$, $\mu $ is the chemical
potential of the ground state, $\psi $ is the ground state of GP equation
and the projection operators $Q$ are given by $Q=1-|\psi \rangle \langle
\psi |.$

The number of noncondensed atoms describes the deviation from condensate
wavefunction and its growth rate characterizes the stability of the
condensate. If the motion of the condensate is stable, the number of
noncondensed atoms grows at most polynomially with time and no fast
depletion of the condensate is expected. In contrast, if the motion of the
condensate is chaotic, the number of noncendensed atoms diverges
exponentially with time and the condesate may be destroyed in a short time.
Therefore the rate of growth of the noncondensed atoms number is similar the
Lyapunov exponent for the divergence of trajectories in phase space for
classical systems \cite{zoller}.

\section{Weak Interactions: Anti-Resonance and Quantum Beating}

In this section, we focus our attention on the case of quantum
anti-resonance, and show how weak interactions between atoms modify the
dynamics of the condensate. Quantum anti-resonace is a single particle
phenomenon characterized by periodic recurrence between two different
states, and its dynamics may be described by Eq. (\ref{G-P}) with parameters 
$g=0$ and $T=2\pi $ \cite{anti-res}.

\subsection{Quantum Beating}

In a non-interacting gas, the energy of each particle oscillates between two
values because of the periodic recurrence of the quantum states. In the
presence of interactions, single particle energy loses its meaning and we
may evaluate the mean energy of each particle 
\begin{equation}
E(t)=\int_{0}^{2\pi }d\theta \psi ^{\ast }\left( -\frac{1}{2}\frac{\partial
^{2}}{\partial \theta ^{2}}+\frac{1}{2}g\left\vert \psi \right\vert
^{2}\right) \psi .  \label{Energy}
\end{equation}

To determine the evolution of the energy, we numerically integrate Eq.(\ref%
{G-P}) over a time span of $100$ kicks, using a split-operator method \cite%
{split}, with the initial wavefunction $\psi $ being the ground state $\psi
\left( \theta ,0\right) =1/\sqrt{2\pi }$. After each kick, the energy $E(t)$
is calculated and plotted as shown in Fig.1.

\begin{figure}[t]
\begin{center}
\vspace*{-0.0cm}
\par
\resizebox *{8cm}{7.5cm}{\includegraphics*{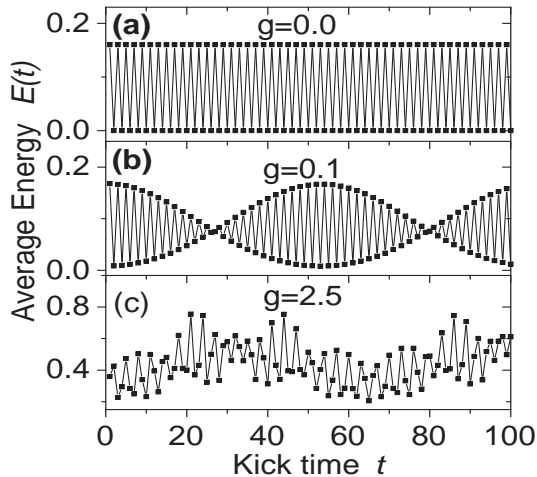}}
\end{center}
\par
\vspace*{-0.0cm}
\caption{Plots of average energy $E(t)$ versus the number of kicks $t$ for
three values of $g$. The kick strength $K=0.8$. }
\label{fig:gp}
\end{figure}

In the case of non-interaction (Fig.1(a) $g=0$), we see that the energy $%
E(t) $ oscillates between two values and the oscillation period is $2T$,
indicating the periodic recurrence between two states (anti-resonance).

\begin{figure}[t]
\begin{center}
\vspace*{-0.0cm}
\par
\resizebox *{7cm}{7.5cm}{\includegraphics*{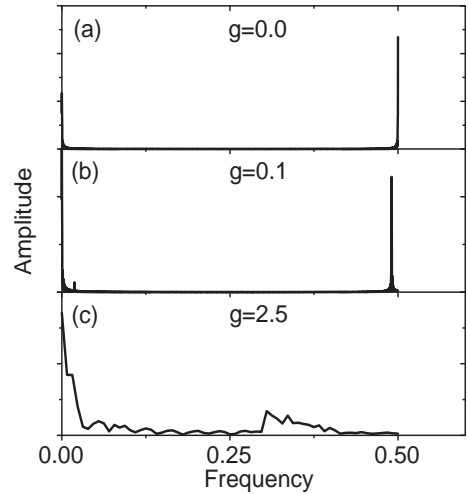}}
\end{center}
\par
\vspace*{-0.0cm}
\caption{Fourier transformation of the energy evolutions in Fig.1. The Unit
for the frequency is $1/T$.}
\label{fig:2}
\end{figure}

The energy oscillation with weak interaction ($g=0.1$) in Fig.1(b) shows a
remarkable difference from that for non-interaction case. We see that the
amplitude of the oscillation decreases gradually to zero and then revives,
similar to the phenomena of beating in classical waves. Clearly, it is the
interactions between atoms in BEC that modulate the energy oscillation and
produce the phenomena of \textit{quantum beating}. As we know from classical
waves, there must be two frequencies, \textit{oscillation} and \textit{beat}%
, to create a beating. These two frequencies are clearly seen in Fig.2 that
is obtained through Fourier Transform of the energy evolutions in Fig.1. For
the non-interaction case (Fig.2(a)), there is only the oscillation frequency 
$f_{osc}=0.5/T$, corresponding to one oscillation in two kicks. The
interactions between atoms develop a new beat frequency $f_{beat}$, as well
as modify the oscillation frequency $f_{osc}$, as shown in Fig.2(b).

\begin{figure}[b]
\vspace*{-0.0cm}
\par
\begin{center}
\resizebox *{8cm}{4cm}{\includegraphics*{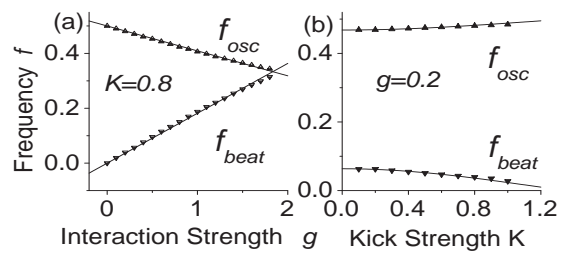}}
\end{center}
\par
\vspace*{-0.0cm}
\caption{Plots of beat and oscillation frequencies versus the interaction
strength (a) and kick strength (b), where the scatters are the results from
numerical simulation using GP equation and lines from analytic expression
Eqs. (\protect\ref{fbeat}) and (\protect\ref{fosci}).}
\label{fig:frequency}
\end{figure}

In Fig. 3, these two frequencies are plotted with respect to different
interacting constant $g$ and kick strength $K$. We see that both beat and
oscillation frequencies vary near linearly with reapect to the interaction
strength $g$. More interestingly, the two frequencies satisfy a conservation
relation 
\begin{equation}
f_{osc}+f_{beat}/2=1/2.  \label{Freq1}
\end{equation}

For a strong interaction [Fig.1(c)], i.e., $g\geq 1.96$, we find that the
energy's evolution demonstrates an irregular pattern, clearly indicating the
collapse of the quasiperiodic motion and the occurrence of instability. The
corresponding Fourier transformation of the energy evolution (Fig.2(c)) has
no sharp peak. This transition to instability will be discussed in Section
IV in details.

\subsection{Spin model}

The phenomena of quantum beating can be understood by considering a two-mode
approximation \cite{liu2} to the GP equation. In this approximation,
condensed atoms can only effectively populate the two lowest second
quantized energy modes. The validity of this two-mode model is justified by
the following facts which are observed in the numerical simulation. First,
the total energy of the condensate is quite small so that we can neglect the
population at high energy modes and keep only the states with quantized
momentums $0$ and $\pm 1$. Second, the total momentum of the condensate is
conserved during the evolution. Therefore the populations of the states with
momentum $\pm 1$ are same if the initial momentum of the condensate is zero
(the ground state).

Here we consider a quantum approach of this two-mode model, which yields an
effective spin Hamiltonian in the mean field approximation. By considering
the conservation of parity we may write the Boson creation operator for the
condensate as 
\begin{equation}
\hat{\psi}^{\dagger }=\frac{1}{\sqrt{2\pi }}\left( \hat{a}^{\dagger }+2\hat{b%
}^{\dagger }\cos \theta \right)  \label{tlwave}
\end{equation}%
where $\hat{a}^{\dagger }$ and $\hat{b}^{\dagger }$ are the creation
operators for the ground state and the first excited states, satisfying the
commutation relation $\left[ \hat{a},\hat{a}^{\dagger }\right] =1$, $\left[ 
\hat{b},\hat{b}^{\dagger }\right] =1$, and particle number conservation $%
\hat{a}^{\dagger }\hat{a}+\hat{b}^{\dagger }\hat{b}=N$.

Substituting Eq. (\ref{tlwave}) into the many-body Hamiltonian of the system 
\begin{eqnarray}
\hat{H} &=&\int_{0}^{2\pi }d\theta \left( \hat{\psi}^{\dagger }\left( -\frac{%
1}{2}\frac{\partial ^{2}}{\partial \theta ^{2}}\right) \hat{\psi}+\frac{g}{2N%
}\hat{\psi}^{\dagger }\hat{\psi}^{\dagger }\hat{\psi}\hat{\psi}\right. 
\notag \\
&&\left. +\hat{\psi}^{\dagger }\left( K\cos \theta \delta _{T}\left(
t\right) \right) \hat{\psi}\right) ,  \label{Ham1}
\end{eqnarray}%
we obtain a quantized two-mode Hamiltonian 
\begin{eqnarray}
\hat{H} &=&-\frac{1}{2}\hat{L}_{z}+\frac{g}{4\pi N}\left( 4\hat{L}_{x}^{2}-%
\frac{1}{2}\left( N-1\right) \hat{L}_{z}+\frac{1}{2}\hat{L}_{z}^{2}\right) 
\notag \\
&&+\sqrt{2}K\hat{L}_{x}\delta _{T}\left( t\right)  \label{Ham2}
\end{eqnarray}%
in terms of the Bloch representation by defining the angular momentum
operators, 
\begin{eqnarray}
\hat{L}_{x} &=&\frac{\hat{a}^{\dagger }\hat{b}+\hat{a}\hat{b}^{\dagger }}{2},
\label{Ang1} \\
\hat{L}_{y} &=&\frac{\hat{a}^{\dagger }\hat{b}-\hat{a}\hat{b}^{\dagger }}{2i}%
,  \notag \\
\hat{L}_{z} &=&\frac{\hat{a}^{\dagger }\hat{a}-\hat{b}^{\dagger }\hat{b}}{2},
\notag
\end{eqnarray}%
where we have discarded all $c$-number terms.

The Heisenberg equations of motion for the three angular momentum operators
reads%
\begin{eqnarray}
\frac{d\hat{L}_{x}}{dt} &=&\frac{\hat{L}_{y}}{2}+\frac{g\left( N-1\right) }{%
4N}\hat{L}_{y}-\frac{g}{8\pi N}\left\{ \hat{L}_{y},\hat{L}_{z}\right\}
\label{Heisen} \\
\frac{d\hat{L}_{y}}{dt} &=&-\frac{\hat{L}_{x}}{2}-\frac{g\left( N-1\right) }{%
8\pi N}\hat{L}_{x}-\frac{7g}{8\pi N}\left\{ \hat{L}_{z},\hat{L}_{x}\right\} 
\notag \\
&&-\sqrt{2}K\hat{L}_{z}\delta _{T}\left( t\right)  \notag \\
\frac{d\hat{L}_{z}}{dt} &=&\frac{g}{\pi N}\left\{ \hat{L}_{y},\hat{L}%
_{x}\right\} +\sqrt{2}K\hat{L}_{y}\delta _{T}\left( t\right)  \notag
\end{eqnarray}%
where $\left\{ \hat{L}_{i},\hat{L}_{j}\right\} =\hat{L}_{i}\hat{L}_{j}+\hat{L%
}_{j}\hat{L}_{i}$.

The mean field equations for the first order expection values $\left\langle 
\hat{L}_{i}\right\rangle $ of the angular momentum operators are obtained by
approximating second order expectation values $\left\langle \hat{L}_{i}\hat{L%
}_{j}\right\rangle $ as products of $\left\langle \hat{L}_{i}\right\rangle $
and $\left\langle \hat{L}_{j}\right\rangle $, that is, $\left\langle \hat{L}%
_{i}\hat{L}_{j}\right\rangle =\left\langle \hat{L}_{i}\right\rangle
\left\langle \hat{L}_{j}\right\rangle $ \cite{anglin}. Defining the
single-particle Bloch vector 
\begin{equation}
\vec{S}=\left( S_{x},S_{y},S_{z}\right) =\frac{2}{N}\left( \left\langle \hat{%
L}_{x}\right\rangle ,\left\langle \hat{L}_{y}\right\rangle ,\left\langle 
\hat{L}_{z}\right\rangle \right) ,  \label{vector}
\end{equation}%
we obtain the nonlinear Bloch equations 
\begin{eqnarray}
\overset{.}{S}_{x} &=&\left( \frac{1}{2}+\frac{g}{8\pi }-\frac{g}{8\pi }%
S_{z}\right) S_{y},  \label{Bloch} \\
\overset{.}{S}_{y} &=&\left( -\frac{1}{2}-\frac{g}{8\pi }-\frac{7g}{8\pi }%
S_{z}\right) S_{x}-\sqrt{2}KS_{z}\delta _{T}\left( t\right) ,  \notag \\
\overset{.}{S}_{z} &=&\frac{g}{\pi }S_{y}S_{x}+\sqrt{2}KS_{y}\delta
_{T}\left( t\right) ,  \notag
\end{eqnarray}%
where we have used $N\gg 1$. The mean field Hamiltonian in the spin
representation reads 
\begin{equation}
\hspace*{-0.5cm}\mathcal{H}=-\frac{S_{z}}{2}+\frac{g}{2\pi }\left( S_{x}^{2}-%
\frac{S_{z}}{4}+\frac{S_{z}^{2}}{8}\right) +\sqrt{2}KS_{x}\delta _{t}(T).
\label{spin}
\end{equation}%
From the definition of the Bloch vector, we see that $S_{z}$ corresponds to
the population difference and $-\arctan (S_{y}/S_{x})$ corresponds to the
relative phase $\alpha $ between the two modes. This Hamiltonian is similar
to a kicked top model \cite{kicktop}, but here the evolution between two
kicks is more complicated.

With the spin model, we can readily study the dynamics of the system. For
the non-interaction case ($g=0$), The Bloch equations (\ref{Bloch}) become%
\begin{eqnarray}
\overset{.}{S}_{x} &=&\frac{1}{2}S_{y},  \label{Bloch2} \\
\overset{.}{S}_{y} &=&-\frac{1}{2}S_{x}-\sqrt{2}KS_{z}\delta _{T}\left(
t\right) ,  \notag \\
\overset{.}{S}_{z} &=&\sqrt{2}KS_{y}\delta _{T}\left( t\right) .  \notag
\end{eqnarray}%
We see that the evolution between two consecutive kicks is simply an angle $%
\pi $ rotation about the $z$ axis, which yields the spin transformation $%
S_{x}\rightarrow -S_{x}$, $S_{y}\rightarrow -S_{y}$. The spin initially
directing to north pole ($\vec{S}=\left( 0,0,1\right) $) stays there for
time duration $T$, then the first kick rotates the spin by an angle $\sqrt{2}%
K$ about the $x$ axis and now $\vec{S}=\left( 0,-\sin \left( \sqrt{2}%
K\right) ,\cos \left( \sqrt{2}K\right) \right) $. The following free
evolution rotates the spin to $\vec{S}=\left( 0,\sin \left( \sqrt{2}K\right)
,\cos \left( \sqrt{2}K\right) \right) $. Then, the second kick will drive
the spin back to north pole through another rotation of $\sqrt{2}K$ about
the $x$ axis. With this the spin's motion is two kick period recurrence and
quantum anti-resonance occurs.

\begin{figure}[t]
\vspace*{-0.0cm}
\par
\begin{center}
\resizebox *{8cm}{4.5cm}{\includegraphics*{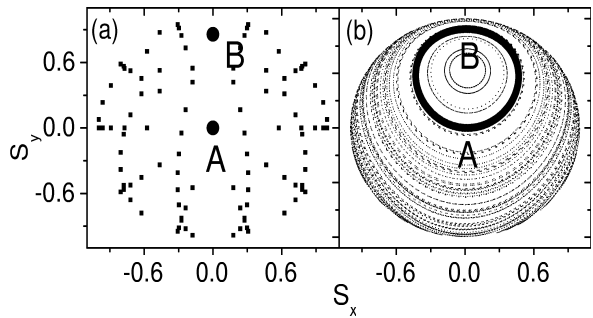}}
\end{center}
\par
\vspace*{-0.0cm}
\caption{Periodic stroboscopic plots of the projection of spin on the $%
S_{z}=0$ plane. The thick line and dots correspond to the orbits with
initial spin $\vec{S}=\left( 0,0,1\right) $. $K=0.8$ (a) $g=0.0$; (b) $g=0.1$%
.}
\label{fig:spin}
\end{figure}

The motion of the spin is more complicate with interactions. The spin
components at $\hat{x}$ and $\hat{y}$ directions can be written as $S_{x}=%
\sqrt{1-S_{z}^{2}}\cos \alpha $ and $S_{y}=-\sqrt{1-S_{z}^{2}}\sin \alpha $
in terms of population difference $S_{z}$ and relative phase $\alpha $,
which yields the relation%
\begin{equation*}
\dot{S}_{x}=\dot{\alpha}S_{y}-\frac{g}{\pi }S_{z}S_{y}\cos ^{2}\alpha 
\end{equation*}%
during the free evolution. Comparing this equation with the Bloch equation (%
\ref{Bloch}), we obtain the equation of motion for the relative phase%
\begin{equation}
\dot{\alpha}=\frac{1}{2}+\frac{g}{8\pi }+\frac{g}{2\pi }\left( \cos 2\alpha +%
\frac{3}{4}\right) S_{z}.  \label{rp}
\end{equation}%
We see the motion between two consecutive kicks is approximately described
by a rotation of $\pi +g(1+3S_{z})/4$ about the $z$ axis. Compared with the
noninteraction case, the mean field interaction leads to an additional phase
shift $g(1+3S_{z})/4$. This phase shift results in a deviation of the spin
from $S_{x}=0$ plane at time $2T^{-}$, i.e., moment just before the second
kick. As a result, the second kick cannot drive the spin back to its initial
position and quantum anti-resonance is absent. However, the phase shift will
be accumulated in future evolution and the spin may reach $S_{x}=0$ plane at
a certain time $mT^{-}$(beat period) when the total accumulated phase shift
is $\pi /2$. Then the next kick will drive spin back north pole by applying
an angle $\sqrt{2}K$ rotation about the $x$ axis.

The above picture is confirmed by our numerical solution of the spin
Hamiltonian with fourth-order Runge-Kutta method \cite{recipe}. In Fig. 4,
we plot the phase portraits of the spin evolution by choosing different
initial conditions of the population difference $S_{z}$ and relative phase $%
\alpha $. Just after each kick three spin components $S_{i}$ ($i=x,y,z$) are
determined and their projections on $S_{z}=0$ plane are plotted. For the
noninteraction case (Fig.4(a)), we see the motion of the spin is an
osillation between the north pole $A$ and another point $B$, indicating the
occurence of quantum anti-resonance. The interaction between atoms changes
the phase portraits drametically (Fig.4(b)). Around the north pole, a fixed
point surrounded by periodic elliptic orbits appears. The two-point
oscillation is shifted slowly and form a continuous and closed orbit,
representing the phenomenon of quantum beating.

\begin{figure}[t]
\vspace*{-0.0cm}
\par
\begin{center}
\resizebox *{8cm}{4cm}{\includegraphics*{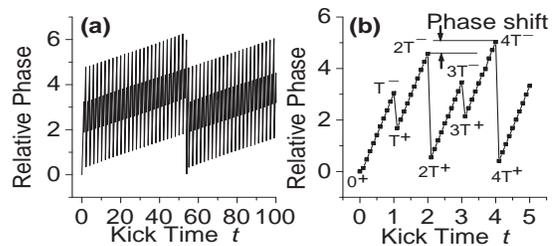}}
\end{center}
\par
\vspace*{-0.0cm}
\caption{(a) Plots of relative phase versus the number of kicks $t$, where $%
g=0.1$, $K=0.8$. (b) Schematic plot of the phase shift. $nT^{-(+)}$
represents the moment just before (after) the $n$th kick.}
\label{fig:classical1}
\end{figure}
\begin{figure}[b]
\vspace*{-0.0cm}
\par
\begin{center}
\resizebox *{8cm}{4cm}{\includegraphics*{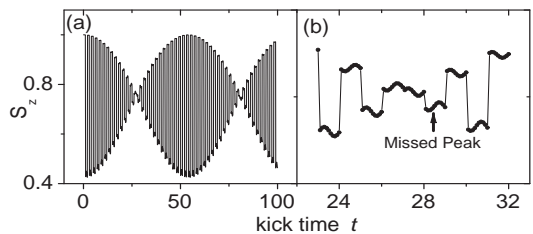}}
\end{center}
\par
\vspace*{-0.0cm}
\caption{(a) Plot of population difference versus the number of kicks $t$,
where $g=0.1$, $K=0.8$. (b) Details of (a) at the middle of a beat period.
The population difference decreases in two consecutive kicks. }
\label{fig:classical2}
\end{figure}

In Fig.5 we see that the relative phase at the moment just before the even
kicks increases almost linearly and reaches $2\pi $ in a beat period. The
slope of the increment reads, $\gamma _{RP}=\left( \alpha \left(
4T^{-}\right) -\alpha \left( 2T^{-}\right) \right) /2$, which can be deduced
analytically. With this and through a lengthy deduction, we obtain an
analytic expression for the beat frequency to first order in $g$, 
\begin{equation}
f_{beat}\approx \frac{g}{4\pi }\left( 1+3\cos \left( \sqrt{2}K\right)
\right) .  \label{fbeat}
\end{equation}

In Fig. 6, we plot the evolution of population difference and the phenomenon
of quantum beating is clearly seen in the spin model. Notice that there is
one peak miss of the oscillation at the middle of one period because the
population difference decreases in two consecutive kicks. Taking account of
this missed peak, we obtain the oscillation frequency%
\begin{equation}
f_{osc}\equiv \frac{N_{total}-N_{miss}}{2N_{total}}=\frac{1}{2}-\frac{1}{2}%
f_{beat},  \label{fosci}
\end{equation}%
where $N_{total}$ and $N_{miss}$ are total and missed numbers of peaks,
respectively.

The analytical expressions Eqs. (\ref{fbeat}) and (\ref{fosci}) of the beat
and oscillation frequencies are in very good agreement with the numerical
results obtained from G-P equation, as shown in Fig.3. Therefore the beating
provides a method to measure interaction strength in an experiment.

\subsection{Anti-Resonance with interactions}

\begin{figure}[b]
\begin{center}
\resizebox*{8cm}{4cm} {\includegraphics*{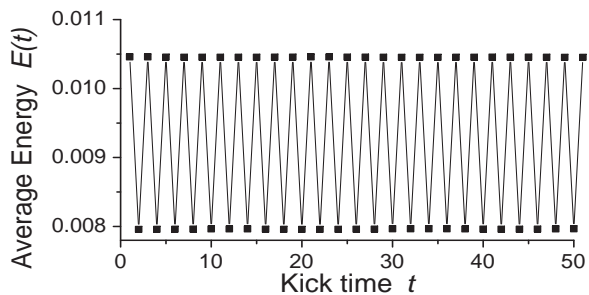}}
\end{center}
\caption{Plots of average energy $E\left( t\right) $ versus the number of
kicks $t$ for $K=0.1$, $g=0.1$. $T_{AR}$ is determined through Eq.(  \protect
\ref{perar}). }
\end{figure}
In the spin model, we see that it is the additional phase shift originating
from weak interactions that destroys the two kick period recurrence of the
anti-resonance and leads to the phenomenon of quantum beating. Therefore we
will still be able to observe the quantum anti-resonance even in the
presence of interactions if the additional phase shift may be compensated.
Actually, the additional phase shift can be cancealed by varying the kick
period $T$ so that the relative phase $\alpha $ only change $\pi $ between
two consecutive kicks. Using Eq. (\ref{rp}), we find the new kick period for
anti-resonance in the presence of interactions may be approximated as%
\begin{equation}
T_{AR}\approx \frac{8\pi ^{2}}{4\pi +g+3g\cos \left( \sqrt{2}K\right) }
\label{perar}
\end{equation}

In Fig. 7, we plot the evolution of the average energy $E\left( t\right) $
with the new kick period $T_{AR}$. We see that the energy oscillates between
two values and the oscillation period is $2T_{AR}$, clearly indicating the
recovery of the anti-resonance.

\subsection{Decoherence due to thermal noise}

\begin{figure}[t]
\begin{center}
\resizebox *{8cm}{8cm} {\includegraphics*{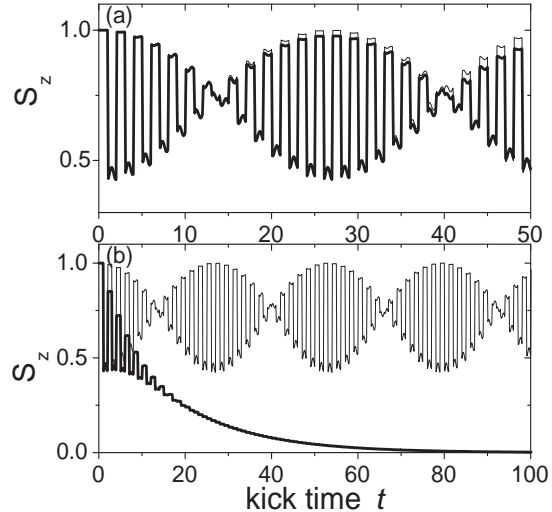}}
\end{center}
\caption{Plots of population difference with respect to the number of kicks
in the presence of decoherence. $K=0.8$, $g=0.2$. Thin lines correspond to $%
\Gamma =0$. Thick lines correspond to (a) $2\protect\pi \Gamma /g=0.001$,
(b) $2\protect\pi \Gamma /g=1$.}
\label{fig:spind}
\end{figure}
In realistic experiments, the decoherence effects always exist. Generally,
decoherence originates in the coupling to a bath of unobserved degree of
freedom, or the interparticle entanglement process \cite{decoh}. The main
source of decoherence in a BEC is the thermal cloud of particles surrounding
the condensate. Thermal particles scattering off the condensate will cause
phase diffusion at a rate proportional to the thermal cloud temperature.
Here we consider a simple model that accounts for the effect of the thermal
noise on the two-mode dynamics by adding a $\tau =1/\Gamma $ transversal
relaxation term \cite{anglin} into the mean-field equations of the motion 
\begin{eqnarray}
\overset{.}{S}_{x} &=&\left( \frac{1}{2}+\frac{g}{8\pi }-\frac{g}{8\pi }%
S_{z}\right) S_{y}-\Gamma S_{x},  \label{deco} \\
\overset{.}{S}_{y} &=&\left( -\frac{1}{2}-\frac{g}{8\pi }-\frac{7g}{8\pi }%
S_{z}\right) S_{x}-\sqrt{2}KS_{z}\delta _{T}\left( t\right) -\Gamma S_{y}, 
\notag \\
\overset{.}{S}_{z} &=&\frac{g}{\pi }S_{y}S_{x}+\sqrt{2}KS_{y}\delta
_{T}\left( t\right) .  \notag
\end{eqnarray}

In Fig.8, we plot the evolution of the population difference $S_{z}$ for
different decoherence constant $\Gamma $. We see that the phenomenon of
quantum beating is destroyed by strong thermal noise (Fig.8(b)), while
survives in weak noise (Fig.8(a)). For large decoherence constant, the
population difference $S_{z}$ decays to $0$ exponentially and the
characteristic time is the just the decoherence time $\tau =1/\Gamma $.
Therefore the decoherence time $\tau $ must be much larger than the beat
period $2\pi /f_{beat}$ to observe the phenomenon of quantum beating, which
yields%
\begin{equation}
2\pi \Gamma /g<<\frac{1}{4\pi }(1+3\cos (\sqrt{2}K)).  \label{deco2}
\end{equation}%
In the case of $K=0.8$, Eq. (\ref{deco2}) gives an estimation $2\pi \Gamma
/g<<0.2$, which agrees with the numerical results shown in Fig.8.

\section{Strong Interactions: Transition to Instability}

\subsection{Characterization of the instability: Bogoliubov excitation}

Tuning the interaction strength still larger means enhancing further the
nonlinearity of the system. From our general understanding of nonlinear
systems, we expect that the solution will be driven towards chaos, in the
sense of exponential sensitivity to initial condition and random evolution
in the temporal domain. The latter character has been clearly displayed by
the irregular pattern of the energy evolution in Fig.1(c). On the other
hand, the onset of instability (or chaotic motion) of the condensate is
accompanied with the rapid proliferation of thermal particles. Within the
formalism of Castin and Dum \cite{castin} described in Section II, the
growth of the number of the noncondensed atom will be exponential, similar
to the exponential divergence of nearby trajectories in phase space of
classical system. The growth rate of the noncondensed atoms is similar to
the Lyapounov exponent, turning from zero to nonzero as instability occurs.

\begin{figure}[t]
\vspace*{-0.0cm}
\par
\begin{center}
\resizebox *{8cm}{4.8cm}{\includegraphics*{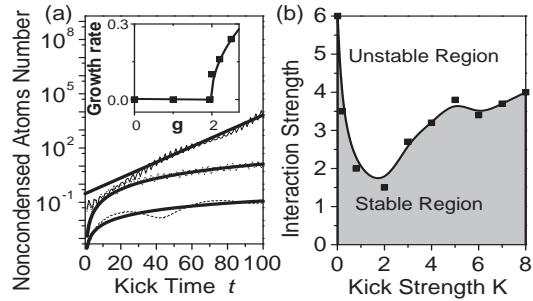}}
\end{center}
\par
\vspace*{-0.0cm}
\caption{(a) Semilog plot of the mean number of noncondensed atoms versus
the number of kicks $t$. The thicker lines are fitting functions. $K=0.8$, $%
g=0.1$ (dashed line, fitting function $0.0003t^{1.3}$), $g=1.5$ (dotted
line, fitting function $0.0011t^{2}$), $g=2.0$ (dash dotted line, fitting
function 0.32exp(0.1t)). The inset shows the interaction dependence of the
growth rate. The scatters are from numerical simulation and the solid line
is the fitting function $0.33{(g-1.96)}^{1/2}$. (b) Phase diagram of the
transition to instability.}
\end{figure}

We numerically integrate Bogoliubov equation (\ref{Bogo}) for the $u_{k}$, $%
v_{k}$ pairs over a time span of 100 kicks, using a split operator method,
parallel to numerical integration of GP equation (\ref{G-P}). The initial
conditions%
\begin{equation}
\left( 
\begin{array}{c}
u_{k}(0) \\ 
v_{k}(0)%
\end{array}%
\right) =\frac{1}{2}\left( 
\begin{array}{c}
\zeta +\zeta ^{-1} \\ 
\zeta -\zeta ^{-1}%
\end{array}%
\right) e^{ik\theta }  \label{ground}
\end{equation}%
for initial ground state wavefunction $\psi (\theta )=1/\sqrt{2\pi }$, are
obtained by diagonalizing the linear operator in Eq.(\ref{Bogo}) \cite%
{castin2}, where $\zeta =\left( \frac{k^{2}/2}{k^{2}/2+2g\left\vert \psi
\right\vert ^{2}}\right) ^{1/4}$.

\begin{figure}[b]
\vspace*{-0.0cm}
\par
\begin{center}
\resizebox *{8cm}{6.14cm}{\includegraphics*{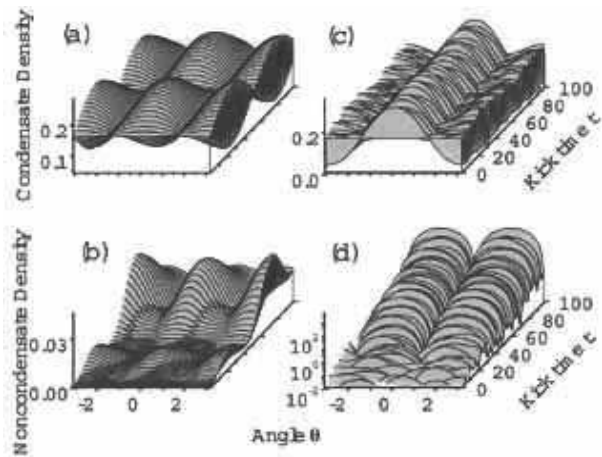}}
\end{center}
\par
\vspace*{-0.0cm}
\caption{Plots of condensate and noncondensate densities, where $K=0.8$.
(a,b) $g=0.1$; (c,d) $g=2.0$. }
\label{fig:wave}
\end{figure}
After each kick the mean number of noncondensed atoms is calculated and
plotted versus time in Fig.9(a). We find that there exists a critical value
for the interaction strength, i.e., $g_{c}=1.96$, above which, the mean
number of noncondensed atoms increases exponentially, indicating the
instability of BEC. Below the critical point, the mean number of
noncondensed atoms increases polynomially. As the nonlinear parameter
crosses over the critical point, the growth rate turns from zero to nonzero,
following a square-root law (inset in Fig.9(a)). This scaling law may be
universal for Bogoliubov excitation as confirmed by recent experiments \cite%
{ketterle}

The critical value of the interaction strength depends on the kick strength.
For very small kick strength, the critical interaction is expected to be
large, because the ground state of the ring-shape BEC with repulsive
interaction is dynamically stable \cite{biao}. For large kick strength, to
induce chaos, the interaction strength must be large enough to compete with
the external kick potential. So, in the parameter plane of $(g,k)$, the
boundary of instability shows a "U" type curve (Fig.9(b)).

Across the critical point, the density profiles of both condensed and
noncondensed atoms change dramatically. In Fig.10, we plot the temporal
evolution of the density distributions of condensed atoms as well as
noncondensed atoms. In the stable regime, the condensate density oscillates
regularly with time and shows clear beating pattern (Fig.10(a)), whereas the
density of the noncondensed atoms grows slowly and shows main peaks around $%
\theta =\pm \pi $ and $0$, besides some small oscillations (Fig.10(b)). In
the unstable regime, the temporal oscillation of the condensate density is
irregular (Fig.10(c)), whereas the density of noncondensed atoms grows
explosively with the main concentration peaks at $\theta =\pm \pi /2$ where
the gradient density of the condensed part is maximum (Fig.10(d)). Moreover,
our numerical explorations show that the $\cos ^{2}\theta $ mode (Fig10.(b))
dominates the density distribution of the noncondensed atoms as the
interaction strength is less than 1.8. Thereafter, the $\sin ^{2}\theta $
mode grows while $\cos ^{2}\theta $ mode decays, and finally $\sin
^{2}\theta $ mode become dominating in the density distribution of
noncondensed atoms above the transition point (Fig10.(d)). Since the density
distribution can be measured in experiment, this effect can be used to
identify the transition to instability.

\subsection{Arnold Diffusion}

\begin{figure}[b]
\begin{center}
\resizebox *{8cm}{14cm}{\includegraphics*{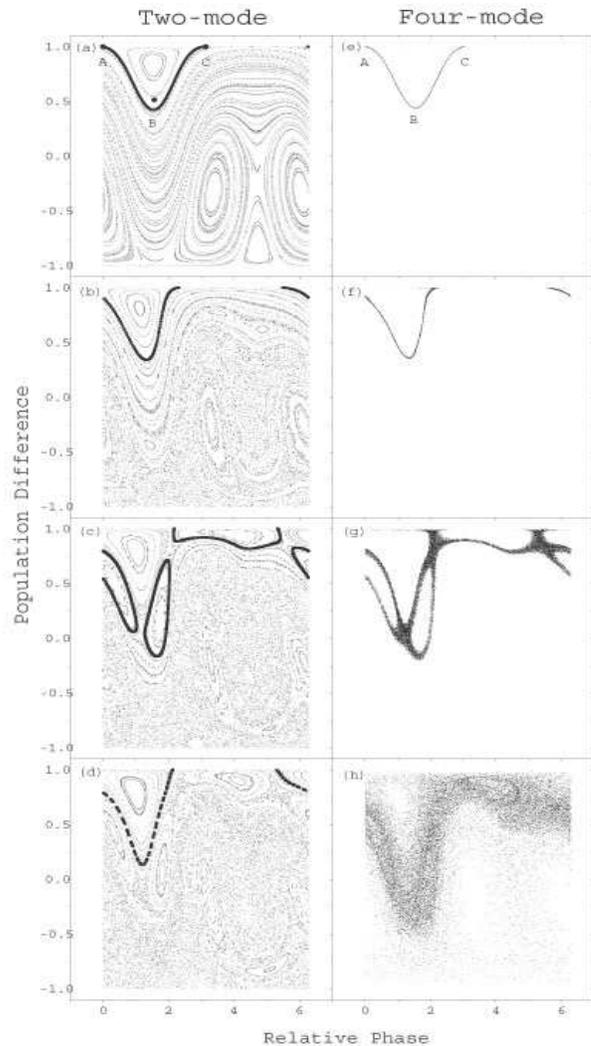}}
\end{center}
\caption{Periodic stroboscopic plots of population difference with respect
to the relative phase between the first two modes. $K=0.8$. (a-d) the
two-mode model, where (a) $g=0.1$, circle dots corresponds to $g=0$; (b) $%
g=1.5$; (c) $g=1.9$; (d) $g=2.0$; The thicker line and larger dots on the
phase portaits represent the trajectories with initial conditions $S_{z}=1,$ 
$\protect\alpha =0$. (e-h) the four-mode approximation, the portait is the
projection on the first two modes of the trajectory with initial condition $%
S_{z}=1,$ $\protect\alpha =0$. (e) $g=0.1$; (f) $g=1.6$; (g) $g=2.2$; (h) $%
g=2.5$.}
\label{fig:classical4}
\end{figure}

We have seen that strong interactions destroy beating solution of the GP
equation and the motion of the condensate become chaotic, characterized by
exponential growth in the number noncondensed atoms. The remaining question
is how the motion of condensate is driven to chaos, that is, the route of
the transition to instability.

The transition to chaos for the motion of the condensate can be clearly seen
in the periodic stroboscopic plots of the trajectories for both two-mode
approximation (Fig.11(a-d)) and four-mode approximation (Fig.11(e-f), it is
exact for the interaction region we consider in Fig.11) to the original GP
equation (\ref{G-P}). The solution oscillates between two points $A$ and $B$
(point $C$ is identical to $A$ in spin model.) for noninteraction case, and
forms closed path in the phase space for the weak interaction (Fig11.(a)).
Note that compared with the two-mode calculation, the effective interaction
strength in the four-mode approximation is rescaled, to give the comparable
pattern in the phase space.

With increasing interaction strength, the stable quasi-periodic orbits in
Fig.11(b) bifurcates into three closed loops (Fig.11(c)) and chaos appears
in the neighborhood of the hyperbolic fixed points. However, diffusion from
one stochastic region to another are still blocked by KAM tori for the
two-mode system. In Fig.11(d), the trajectory with initial condition $%
S_{z}=1,\alpha =0$ is closer to the chaotic region.

The above discussion is based on two-mode approximation, actually, the
solution is coupled with other modes of higher energy states. For small
interaction, this coupling is negligible and the four-mode simulation gives
the same results as seen in Fig.11(a,b,e,f). For large interaction, this
coupling is important and our system is essentially high-dimensional ($d>2$%
). One important character of a high-dimensional dynamical system is that
KAM tori ($d$-dimension) can not seperate phase space ($2d$-dimension) and
the whole chaotic region is interconnected. If a trajectory lies in a
chaotic region it can circumvent KAM tori and diffuse to higher energy
states through Arnold diffusion. This process is clearly demonstrated in
Fig.11(g,h). We see that the trajectory diffuses along the separatrix
layers, circumvents the KAM tori, and finally spreads over whole phase space
(Fig.11(h), $g=2.5$). We also calculate the diffusion coefficient 
\begin{equation}
D_{E}=\frac{2}{J(J-1)}\sum_{m>n}\frac{\left\vert E_{m}-E_{n}\right\vert ^{2}%
}{T\left( m-n\right) },  \label{diffu}
\end{equation}%
where $E_{m}$ is the energy after the mth kick, $J$ is the total number of
kicks. For $g=2.2$ and $g=2.5$, the diffusion rates are $7.2\times 10^{-11}$
and $1.5\times 10^{-9}$ respectively.

Arnold diffusion allows the state to diffuse into higher energy states,
which destroys quasiperiodic motion of the quantum beating and leads to the
transition to instability. As Arnold diffusion occurs, the motion of the
condensate becomes unstable and the number of the noncondensed atoms grow
exponentially, as we have seen in above discussion.

Arnold diffusion is a general property of the nonlinear Schr\"{o}dinger
equation in the presence of strong interactions. However, it may be hard to
observe the whole process of Arnold diffusion in realistic BEC experiments
because of the limit number of atoms ($10^{6}$). As Arnold diffusion occurs,
the instability of the condensate leads to the exponential growth of thermal
atoms which destroy the condensate, as well as invaildate the GP equation (%
\ref{G-P}) in a short time; while clear signature of the whole Arnold
diffusion process may only be observed in a realtive long period. On the
other hand, Arnold diffusion may be observed in the context of nonlinear
optics, where the GP equation (\ref{G-P}) describes the propagation of
photons. The number of photons is very large and the interactions between
them are very weak, therefore it is possible to have a long diffusion
process without invalidating the GP equation (\ref{G-P}).

\subsection{Dynamical localized states}

Although the above discussions have been focused on a periodic state of
anti-resonance, the transition to instability due to strong interactions
also follows a similar path for a dynamically localized state. The only
difference is that we start out with a quasiperiodic rather than periodic
motion in the absence of interaction. This means that it will generally be
easier to induce instability but still requires a finite strength of
interaction.
\begin{figure}[!t]
\vspace*{-0.0cm}
\par
\begin{center}
\resizebox *{8cm}{4cm}{\includegraphics*{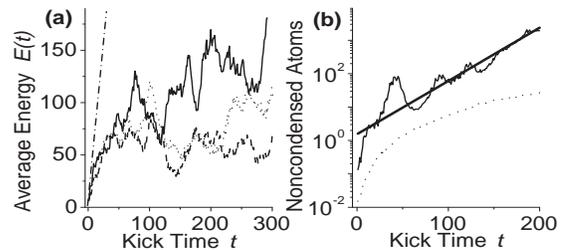}}
\end{center}
\par
\vspace*{-0.0cm}
\caption{Nonlinear effects on dynamically localized states. $K=5$, $T=1$.
(a) Plots of average energy $E(t)$ versus the number of kicks $t$, where
dash dotted line corresponds to the classical diffusion. $g=0$ (dash), $g=1$
(dot), $g=5$ (solid). (b) Semilog plot of the mean number of noncondensed
atoms versus the number of kicks $t$. $g=1$ (dot), $g=5$ (solid).}
\label{fig:dl}
\end{figure}

In Fig.12, we show the nonlinear effect on a dynamically localized state at $%
K=5$ and $T=1$. For weak interactions ($g=1$) the motion is quasiperiodic
with slow growth in the number of noncondensed atoms. Strong interaction ($%
g=5$) destroys the quasiperiodic motion and leads to diffusive growth of
energy, accompanied with exponential growth of noncondensed atoms that
clearly indicates the instability of the BEC. Notice that the rate of growth
in energy is much slower than the classical diffusion rate, which means that
chaos brought back by interaction in this quantum system is still much
weaker than pure classical chaos.

\section{Conclusions}

We have investigated the complex dynamics of a periodically kicked
Bose-Einstein condensate that is considered as a nonlinear generalization of
the quantum kicked rotor. We demonstrate the transition from the
anti-resonance to the quantum beating and then to instability with
increasing many-body interactions, and reveal their underlying physical
mechanism. The stable quasiperiodic motions for weak interactions, such as
anti-resonace and quantum beating, have been studied by mapping the
nonlinear Schr\"{o}dinger equation to a spin model. The transition to
instability has been characterized using the growth rate of the noncondensed
atoms number, which is polynomial for stable motion and exponential for
chaotic motion of the condensate.

Finally, we emphasize that the results obtained in the paper are not limited
to BEC and can be directly applied to other systems whose dynamics are
governed by the nonlinear Schr\"{o}dinger equation.

\begin{acknowledgments}
We acknowledge the support from the NSF, the R. A. Welch foundation, MGR
acknowledges supports from Sid W. Richardson Foundation, JL acknowledges
supports from NSFC(10474008) and CAEP Foundation.
\end{acknowledgments}

\end{document}